\documentclass[10pt]{article}

\def\al{\alpha}
\def\be{\beta}
\def\ga{\gamma}
   
\def\phi{\varphi}
\def\la{\lambda}

\def\si{\sigma}
\def\ze{\zeta}

\def\E{{\mathcal E}}
\def\X{{\mathcal X}}

\def\pd{\partial}
\def\d{{\rm d}}       

\def\~#1{\widetilde #1}

\def\beq{\begin{equation}}
\def\eeq{\end{equation}}
\def\lb{\label}

\def \sy {symmetry}
\def \sys {symmetries}
\def \eq {equation}
\def\={\,=\,}
\def\q{\quad}
\def\d{{\rm d}}
\def\sk{\smallskip}
\def\bk{\bigskip}
\def\pd{\partial}
\def\vf{vector field}
\def\so{solution}

\def\ni{\noindent}

\def\EOE{\hfill{$\triangle$}}

\def\.#1{\dot #1}
\def\^#1{\widehat #1}

\date{}

\begin{document}

\title{Generalized notions of symmetry of ODE's and reduction procedures}

\author{
   Giampaolo Cicogna\thanks{Email: cicogna@df.unipi.it} 
   \\~\\
Dipartimento di Fisica ``E.Fermi'' dell'Universit\`a di Pisa\\
  and  Istituto Nazionale di Fisica Nucleare, Sez. di Pisa \\~\\
Largo B. Pontecorvo 3, Ed. B-C, I-56127, Pisa, Italy  }

\maketitle

\ni{\bf Abstract}

\ni
This paper describes the notion of  $\sigma$-symmetry,
which extends the one of $\lambda$-symmetry, and its application to 
reduction procedures of systems of ordinary differential equations and of dynamical systems as well.
We also consider  orbital symmetries, which give rise 
to a different form of reduction of  dynamical systems. Finally, 
we discuss how dynamical systems can be transformed into  higher-order ordinary differential equations, 
and how these symmetry properties of the dynamical systems can be 
transferred into reduction properties of the corresponding ordinary differential equations. 
Many examples  illustrate the various situations. 

\bigskip \ni

\bigskip \ni

\bigskip\ni
{\bf Talk given at the ICNAAM Conference, Kos (Greece), Sept. 2012. Based on joint work with G. Gaeta and S. Walcher}

\medskip 
\section*{Introduction} 

It is well known that if an ordinary differential \eq\ (ODE)
of order $q>1$ admits a Lie point-\sy ,
then the order of the \eq\ can be lowered by \emph{one} (\emph{two} in 
some cases, e.g. when the \eq\ comes from a variational problem), see 
e.g. \cite{Ovs,Olv,Ste,CRC,BA}.

It is also known that the same is true even if the \eq\  admits a 
$\la$-\sy , 
a notion which has been introduced by C. Muriel and J. Romero in 2001 
\cite{MR1,MR2} and which has received a number of applications and extensions 
(see e.g. \cite{Gtw} with references therein, and \cite{Sig13} 
for a more recent contribution).

We have recently further extended this result \cite{Sprol,CGW,SDS}. 
Let us fix our 
notations. We will always denote by $t$ the independent variable, in order 
to unify the notations, as  a large part of this paper 
will be concerned with dynamical systems, where time $t$ is typically the 
independent variable. The ODE will be denoted by
\[\E\=\E\big(t,u^{(k)}(t)\big)\=0\q\q \big(u^{(k)}(t)=\d^ku/\d t^k \ ,\ 
k=0,\ldots,q\big)\]
and the generators of Lie point-\sys\ will be written in the form
\[X\=\phi(t,u)\frac{\pd}{\pd u}+\tau(t,u)\frac{\pd}{\pd t}\ .\]
According to a by now standard abuse of language, we will denote by $X$ 
both the \sy\ and its Lie generator.

We will consider, instead of a {\it 
single} \vf\ $X$, a set $\X$ of $s>1$ \vf s $X_\al$ in involution
\beq\lb{inv} [X_\al,X_\be]\=\nu_{\al \be \ga }X_\ga 
\q\q (\al,\be,\ga=1,\ldots,s) \eeq
together with a system of ODE's $\E_a=0$, $a=1,\ldots,n$.
This leads to the introduction of the notion of ``combined'' 
\emph{joint-$\la$-\sys }, 
or \emph{$\si$-\sys\ } for short. The precise definition and its 
application to the reduction of
systems of ODE's will be given in the next Section. Using the same idea, 
we will show (Sect. 2) that also dynamical systems (DS), i.e. systems of 
first-order ODE's, can be suitably reduced when they admit a $\si$-\sy .
In Sect. 3, we include the case of \emph{orbital} \sys , which give rise 
to a different form of reduction of  DS. Finally, in Sect. 
4, we discuss how DS can be transformed into a higher-order ODE, and how 
these \sy\ properties of the DS can be transferred into reduction 
properties of the corresponding ODE. Several new examples will illustrate the 
various situations. All the objects (functions, \vf s) considered in this
paper are assumed to  be smooth enough.

The presence of $\si$-\sys\ admits interesting geometrical interpretations 
and algebraic aspects: for a full discussion of these arguments and several other 
details we refer to \cite{Sprol,CGW,SDS} and references therein.

This is a full paper presented within ICNAAM 2012; a very short and 
preliminary sketch of part of these results can be found in the  Enlarged 
Abstracts of the Conference Proceedings \cite{EnAb}. 

\section{Basic definitions and reduction of ODE's}

First of all, we need the two following definitions.

\sk\ni
{\bf Definitions}

\ni
i) \emph {Given $n>1$ variables $u\equiv \{u^a(t)\}$, $(a=1,\ldots,n)$, and
$s>1$ \vf s  $\X\equiv \{X_\al\}$, $(\al=1,\ldots,s)$, 
a $\si$-prolongation is a  
deformed prolongation rule which involves  
a given $s\times s$ matrix $\si=\si(t,u,\dot u)$:
the first $\si$-prolongation $Y_\al^{[1]}$ of 
$X_\al=\phi_\al\cdot\nabla_u+\tau_\al\pd/\pd t$  is defined by
\[Y_\al^{[1]}:=X_\al^{[1],\si}\=X_\al^{[1]}+ \si_{\al\be}(\phi_\be^a-\dot 
u^a\tau_\be)\frac{\pd}{\pd \dot u^a} \]  
where $X_\al^{[1]}$ is the first standard prolongation. 
Higher order prolongations $Y_\al^{[k]}$ can be easily obtained by 
recursion.}
 
\bk\ni
ii)  \emph{A system of $n$ ODE's $\E\equiv \{\E_a\big(t,u^{(k)}(t)\big)\}=0$  
for the $n$ variables $u(t)$, of order $q>1$, is $\si$-symmetric under the 
set $\X$  if
%\vskip-.2cm
\[Y_\al^{[q]}\E|_{\E=0}\=0\]
%\vskip-.2cm
i.e. if $\E$ is invariant under the $\si$-prolongations $Y_\al^{[q]}$ of 
all the $X_\al$.}

\sk

It can be remarked that the case $s=1$ would correspond to $\la$-\sys .

Based on the above definitions, we can state the following result.

\sk\sk\ni
{\bf Theorem 1.} \emph {Let  a system of $n$ ODE's $\E=0$ of order $q>1$ 
be $\si$-symmetric under a set $\X$ of \vf s $X_\al\, (\al=1,\ldots,s>1)$ 
in involution with constant rank $r$ ($r\le s;\, r\le n$); if the 
involution relations are preserved in their $q$-th $\si$-prolongations 
$Y^{[q]}_\al$, then -- under standard regularity and nondegeneracy 
conditions -- the order of $r$ ODE's can be lowered by 
one. This is obtained in terms of some $r$ new variables $\eta_\al$ which 
are invariant under the $1^{st}$ $\si$-prolongations $Y^{[1]}_\al$. }

\sk\ni
{\it Sketch of the proof.} The main ingredient of the proof is the 
following  {\it completely algebraic} result, which holds for general 
\vf s
$X_\al=\phi_\al\cdot\nabla_u+\tau_\al\pd/\pd t$
\beq\lb{DtY} 
[D_t,Y_\al^{[k+1]}]\=-\si_{\al\be}Y^{[k]}_\be+(D_t\tau_\al+
\si_{\al\be}\tau_\be)D_t\eeq
where $D_t$ is the total derivative,
and its consequence 
\beq\lb{IDP}Y_\al^{[k+1]}\,\frac{D_t\ze_1^{[k]}}{D_t\ze_2^{[k]}}\=0\eeq 
where  
$\ze_i^{[k]}$ is any $k$-order differential invariant  under $Y_\al^{[k]}$.
Assume for simplicity (but the general result holds in general) 
that the   $X_\al$ are \emph{vertical} \vf s,  i.e. that $\tau_\al=0$:
then, the time $t$ is a common invariant under all 
the  $X_\al$. Assume also, for the moment, that $n=r$. Then, no 
other variable is admitted with this property.
Considering the first $\si$-prolonged \vf s $Y_\al^{[1]}$,
there exist, according to Frobenius theorem,  exactly $n$ common 
differential 
invariants of order $1$ under $Y_\al^{[1]}$. Let us denote these by 
$\eta_\al\, (\al=1,\ldots, r=n)$. Using (\ref{IDP}) with $k=1$, 
choosing as $\ze_1$ any of these $\eta_\al$ 
and $\ze_2=t$, we deduce that $D_t\eta_\al=\dot\eta_\al$ are second-order  
differential quantities which are common invariants under the second 
$\si$-prolongation $Y_\al^{[2]}$, and so on. This is called 
{\it invariance by differentiation property}.
The $\si$-invariance of the system $\E=0$, then 
implies that all the \eq s of this system must contain, apart from $t$, 
only the common invariant variables  with their derivatives.  Choosing 
$\eta_\al$ as new variables, the \eq s of our system thus become \eq s of 
order 
$q-1$. If instead $n>r$, then, still thanks to Frobenius theorem, there 
are, in addition to $t$, other $(n-r)$ variables $w_j\ 
(j=1,\ldots,n-r)$ of order zero  which are common invariants  under 
$X_\al$. Therefore, thanks to (\ref{IDP}), also $\dot w_j$ are $(n-r)$ common
invariants under the first 
$\si$-prolongation $Y_\al^{[1]}$, in addition to other $r$ invariants 
$\eta_\al$, and so on. In other words, starting from the 
invariants $w_j$ and $\eta_\al$, one obtains all higher-order 
differential invariants. As before, our system  must be written in terms 
of these invariant quantities; then   the 
system of ODE's can be split into a subsystem of $r$ \eq s   
of order $q-1$ in the variables $t$ and $\eta_\al$, 
and another system of $n-r$ \eq s of order $q$. 
\hfill$\bullet$

\sk\ni
{\it Example 1.}
Consider the system of ODE's (in the examples we will usually write  
as $u_1,u_2,\ldots$ instead of $u^a$ to avoid confusion, and $\dot u_1=\d u_1/\d t$, etc.)
\begin{equation}
\left\{\displaystyle \begin{array}{ll}
 \stackrel{\ldots}{u}_1\=t\ddot u_2+t\dot u_2+2\dot u_2+u_2+h_1(t,u)\\
\ddot u_2\=\dot u_1-\dot u_2+h_2(t,u)\\
\ddot u_3\=u_2+t\dot u_2+h_3(t,u)
\end{array}\right.
\end{equation}
where $h_a$ are arbitrary functions of $t$ and of the quantities 
$u_1-u_2-u_3,u_1-u_2-\dot u_1+tu_2,u_1-u_2-\dot u_2$.
For generic $h_1,h_2,h_3$ there is no standard Lie \sy\  for this system, 
but 
it is $\si$-symmetric under the \vf s (then $n=3, r=2$)
\[X_1\=\frac{\pd}{\pd u_1} + \frac{\pd}{\pd u_2}\q,\q X_2\=\frac{\pd}{\pd 
u_1} +\frac{\pd}{\pd u_3}\]
  with    
\[\si\=\pmatrix { 0 & t\cr 1 & 0 }\ .\]
The first $\si$-prolongations are
%\vskip-.2cm
\[Y_1^{[1]}\=X_1+t\frac{\pd} {\pd \dot u_1} +t\frac{\pd} {\pd \dot u_3} 
\q,\q\
 Y^{[1]}_2\=X_2+\frac{\pd} {\pd \dot u_1} + \frac{\pd} {\pd \dot u_2} \ .\]
%\vskip-.2cm
In the new $\si$-\sy\ adapted variables
%\vskip-.2cm
$w=u_1-u_2-u_3,\, \eta_1=u_1-u_2-\dot u_1+tu_2 ,\, \eta_2= u_1-u_2-\dot 
u_2$
%\vskip-.2cm
the above \eq s become, in agreement with Theorem 1,
\[\ddot \eta_1=-\dot\eta_1+\dot\eta_2+h_1(\eta_1,\eta_2,w)\ ,\
\dot \eta_2=-h_2(\eta_1,\eta_2,w)\ ,\ \ddot 
w=\dot\eta_2+\dot\eta_1-h_3(\eta_1,\eta_2,w)
\]\EOE

\sk
It can be observed that if one of the \eq s of the system of ODE's is of
order $1$ and this is 
lowered according to Theorem 1, then one is left with an 
\emph{algebraic} \eq\ for the variables $t$ and $\eta_\al$. This happens 
for 
instance if in Example  above one of the \eq s is replaced by
\[ \dot u_1\=u_1-u_2+tu_2+h_0(t,u)\]
which is reduced to
\[\eta_1+h_0(\eta_1,\eta_2,w)\=0\ .\] 
Notice that this algebraic \eq\ is actually a first-order differential 
\eq\ for the initial variables $u^a$ (the presence of an ``auxiliary'' 
first-order differential \eq\ is indeed standard in $\la$-type \sys ).

This remark introduces the special and specially  interesting case of 
dynamical 
systems, which will be considered in detail in the next sections.

\section{Reduction of Dynamical Systems}

Dynamical systems are systems of first-order time-evolution
differential \eq s of the form
\[\dot u^a\=f^a(t,u)\q\q a=1,\ldots,n\]
It is not too restrictive to consider \emph{autonomous} DS, 
and  vertical  \vf s   with $\phi_\al$ 
\emph{independent of time}, i.e. 
\beq\lb{ass}\dot u=f(u) \q\q\ X_\al\=\phi_\al^a\frac{\pd}{\pd u^a}\equiv 
\phi_\al\cdot\nabla_u\eeq

Given a DS, the $\si$-determining \eq s, i.e. the \eq s giving the 
conditions for the DS to be invariant under the first $\si$-prolongations 
$Y_\al^{[1]}$ of $X_\al$, when 
restricted to the \so \ manifold of the DS, take the particularly simple 
form
%\beq \lb{sideq1} [\phi_\al,f]\=\si_{\al\be}\phi_\be \eeq
\beq\lb{sideq} [X_\al,F]\=\si_{\al\be}X_\be \q\q
(\al,\be=1,\ldots,s)\eeq
having introduced the ``dynamical'' \vf
\[F\=f\cdot\nabla_u \ .\]

In particular, the restriction to the \so \ manifold of the DS $\dot 
u=f(u)$, implies 
that $\si$ may be chosen as a function of $t,u$ only, indeed 
$\si\big(t,u,f(t,u)\big)=\overline{\si}(t,u)$. From (\ref{sideq}), one
may  directly recover for this case the invariance by differentiation property: indeed,
if $w_j$ satisfies $X_\al w_j=0$, then
\[X_\al(D_tw_j)=X_\al(f\cdot\nabla_u)w_j=X_\al F\,w_j=
(FX_\al+\si_{\al\be}X_\be)w_j=0  \]
i.e. $D_tw_j$ is also invariant under all the $X_\al$.

\sk
As well known, given a set $\X$ of \vf s in involution, it is not granted 
in general that their prolongations are still in involution (see \cite{Sprol,SDS} 
for a discussion and some examples on this point). However, in the case of 
DS, we have the following useful result (in the following, we will simply 
write $Y_\al$ instead of $Y_\al^{[1]}$):

\sk\sk\ni
{\bf Lemma}. \emph{Let a DS satisfy (\ref{sideq}) with a set of \vf s
$X_\al$ in  involution. Then, restricting to the \so \ manifold of the
DS, the first 
$\si$-prolonged \vf s $Y_\al$ satisfy the same involution property.}

\sk\ni
{\tt Proof}. We first have
\[Y_\al\=X_\al+\big(D_t\phi^a_\al+\si_{\al\be}\phi^a_\be\big)
\frac{\pd}{\pd \dot u^a}=
X_\al+X_\al \,f\nabla_{\dot u}\]
thanks to (\ref{sideq}). Then
\[[Y_\al,Y_\be]\=[X_\al,X_\be]+\nu_{\al\be\ga}X_\ga f\nabla_{\dot u}
=\nu_{\al\be\ga}Y_\ga\]
using the involution properties of $X_\al$.
\hfill$\bullet$

\sk
Then we have:

\sk\ni
{\bf Theorem 2.} \emph{In the above simplifying assumptions (\ref{ass}),
let  a DS be 
$\si$-symmetric under a set $\X$ of \vf s $X_\al\, (\al=1,\ldots,s>~1)$ in 
involution, with rank $r<n$; then the DS can be locally  
reduced  to a DS involving  $n-r$  variables $w_j$:
\[\dot w_j\=W_j(w)\]
plus a system of $r$ ``reconstruction \eq s " depending on the \so s of
the  reduced system.} 

\sk\sk
As for the above case of general ODE's, the proof is based on the 
introduction of $n$ \sy -adapted variables: precisely of $(n-r)$ variables 
$w_j$ which are 
the entries of the reduced DS and are common invariants under 
$X_\al$, and of $r$ first-order differential $Y_\al$-invariants $\eta_\al$. 

It can be noticed that this reduction to a $n-r$-dimensional 
DS holds exactly as in the case of standard (exact) \sys . See also 
\cite{MRVi} for the case of DS admitting $\la$-\sys .

\sk\sk\ni
{\it Example 2.} This is a very trivial example, given to provide a clear 
illustration of the procedure. The DS
\[
\left\{\displaystyle \begin{array}{ll}
\dot u_1\=h_1(u_1,u_2,u_3)+g_1(u_1-u_3)\\
\dot u_2\=h_2(u_1,u_2,u_3)+g_2(u_1-u_3)\\ 
\dot u_3\=h_1(u_1,u_2,u_3)+g_3(u_1-u_3)
\end{array}\right.
\]
where $h_a,\,g_a$ are arbitrary functions of the indicated
arguments, admits the two \vf s
\[X_1\=\pd/\pd u_1+\pd/\pd u_3 \q ,\q X_2\=\pd/\pd u_2\]
as $\si$-\sy , as can be easily verified, with 
\[\si\=\pmatrix{ \pd h_1/\pd u_1+\pd h_1/\pd u_3 & \pd h_2/\pd u_1+
\pd h_2/\pd u_3\cr \pd h_1/\pd u_2 & \pd h_2/\pd u_2 } \ .\]
In terms of the symmetry-adapted 
variables $w=u_1-u_3$, $\eta_1=\dot u_1-h_1\, ,\,\eta_2=\dot u_2-h_2$, 
the DS becomes
\[\dot w=g_1(w)-g_3(w)\ ,\ \eta_1=g_1(w)\ ,\ \eta_2=g_2(w)\]
where the first \eq\ is the reduced system and the other two the
reconstruction \eq s. \EOE

\sk\sk
In the following, it will be convenient to rewrite \eq\ (\ref{sideq}) in 
the more transparent form, with evident notations,
\beq\lb{sideq1} [\phi_\al,f]\=\si_{\al\be}\phi_\be \q\q 
(\al,\be=1,\ldots,s)\eeq
An important property of $\si$-symmetric DS is given by the following
proposition, which can be easily verified, using (\ref{sideq}) 
(or (\ref{sideq1})):

\sk\ni
{\bf Proposition 1}. \emph{Let $\dot u=f$ be a DS admitting a
set $\X$ of \vf s $X_\al=\phi_\al\cdot\nabla_u$ in involution as a {\rm
standard}
\sy . Then, for any choice of $s$ functions $\mu_\al(u)$, the new DS
\[\dot u\=f^*:=f+\sum_{\al=1}^s \mu_\al \phi_\al\]
admits the set $\X$ as $\si$-\sy , where $\si$ is given by}
\[  \si_{\al\be} \= \ X_\al(\mu_\be)  + \mu_\ga\nu_{\al\ga\be}\ .
\]

\sk
For a partial converse of this result, see \cite{CGW}. The above proposition 
is clearly useful for constructing explicit examples  of $\si$-symmetric
DS (it is known that, given a DS, it may be very difficult to determine 
its $\si$-\sys , because the $\si$-determining 
\eq s are in general differential functional \eq s: see \cite{Sprol,SDS} for a 
discussion on this aspect).

\sk\sk\sk\ni
{\it Example 3.} 
As a special case of the above Proposition, consider $f=Au$ for some 
matrix $A$; then obviously $\phi_\al=B_\al u$, with $B_\al$ matrices such that 
$[A,B_\al]=0$, provide standard symmetry vector fields $X_\al$ for 
$\dot u=Au$. 
These matrices will satisfy $[B_\al,B_\be]=c_{\al\be\ga}B_\ga$ and hence 
$[X_\al,X_\be]=-c_{\al\be\ga}X_\ga$ and the vector 
fields $X_\al$ provide a $\si$-symmetry for the  DS (we can take $B_0=A$)
\[\dot u\= A\,u\ +\ \sum_{\al=0}^s \mu_\al(u)\,B_\al u\ .\]
for any functions $\mu_\al$. As a concrete example, consider the DS
\[\dot u_1\=u_1-u_2\q ,\q \dot u_2\=-u_1+u_2\q ,\q \dot u_3\= a u_3\]
where $a$ is any constant: it admits the two standard \sys
\[X_1\=u_1\frac{\pd}{\pd u_1}+u_2\frac{\pd}{\pd u_2}+
u_3\frac{\pd}{\pd u_3} \ ,\ 
X_2\=u_2\frac{\pd}{\pd u_1}+u_1\frac{\pd}{\pd u_2}\]
Using Prop.1 with $\mu_1=u_1,\,\mu_2=u_3$, we obtain the new DS
\[
\left\{\displaystyle \begin{array}{ll}
\dot u_1\=u_1-u_2+u_1^2+u_1u_2\\
\dot u_2\=-u_1+u_2+u_1u_2+u_1u_3\\ \dot u_3\=a u_3+u_1u_3
\end{array}\right.
\]
which admits the above \vf s as $\si$-\sy . Accordingly (here $n-r=1$), 
we get  in terms of the common invariant variable 
$w=(u_1^2-u_2^2)/u_3^2$ the reduced \eq\ $\dot w=2(1-a)w$.\EOE

\section{Orbital symmetries}

If in the \eq\ (\ref{sideq}) (or (\ref{sideq1}))
the indices $\al,\be$ run from $0$ to $s$,  and some $\si_{\al 
0}:=\theta_\al\not=0$, i.e. if
\beq\lb{orsy} [\phi_\al,f]\=\theta_\al f+\si_{\al\be}\phi_\be \q\q 
(\al,\be=1,\ldots,s)\eeq
the case of  \emph{orbital} $\si$-symmetries  is included.

Let us recall that in the case of \emph{proper} orbital \sys\ (i.e. when 
$\si_{\al\be}=0$, some $\theta_\al\not=0$) we have:

\sk\ni
{\bf Proposition 2.} \emph{If a $n$-dimensional DS $\dot u=f(u)$ admits an 
involutive set of $s\ge 1$ 
\vf s $\X\equiv\{X_\al\}$  as a (proper) orbital \sy , then: 
i) $X_\al$ map \so\ orbits into \so\ orbits;
ii) there is a scalar nonzero function $\rho(u)$ such that the DS
\[\dot u=\rho\,f(u)\] 
is  standardly symmetric under $X_\al$;
iii) the DS $\dot u=\rho\,f(u)$ is \emph{orbitally equivalent} to 
$\dot u=f(u)$, i.e. the two DS have the same \so s orbits and the same 
constants of motion.
The initial DS can then ``orbitally reduced'', i.e. there are $n-r$ 
variables $w_j$ (where $r<n$ is the rank of $\X$), invariant under $X_\al$, and a 
nonzero scalar function $\omega(u)$ such that
\beq\lb{rw} \dot w_j\=\omega(u)\,W_j(w)\eeq
i.e. we get a reduction ``up to a common scalar factor". }

\sk\sk
In the general case of   orbital-$\si$-\sys\ (\ref{orsy}) we have 
essentially the same result:

\sk\ni
{\bf Theorem 3.} \emph{In the simplifying assumptions as above, if a DS 
admits an involutive set $\X$ of orbital $\si$-\sys ,
then the DS can be orbitally reduced as in Proposition 2.}

\sk
As is clear from (\ref{rw}), if we have at least two variables $w_j$, say 
$w_1,w_2$, we can obtain from  (\ref{rw}) a reduced \eq\ of the form
\[\frac{\d w_1}{d w_2}\=\Psi(w_1,w_2)\ .\]

\sk\ni
{\it Example 4.}  Let us use in this case for simplicity the notations 
$u\equiv(x,y,z)$ and $r^2=x^2+y^2,\,\theta=\arctan(y/x)$. Consider the DS 
%\vskip-.1cm
\[
\left\{\displaystyle \begin{array}{ll}
\dot x\=h_1(x,y,z)x+h_2(x,y,z)y\\
\dot y\=h_1(x,y,z)y-h_2(x,y,z)x\\ \dot z\=h_3(x,y,z)z
\end{array}\right.
\]
%\vskip-.1cm
and the rotation \vf\
$X=y\pd/\pd x-x\pd/\pd y=\pd/\pd\theta$; then $w_1=r,\,w_2=z$. We 
distinguish the following cases:

\sk\ni
a) all the $h_i$ are functions of $r^2=x^2+y^2$ and $z$ only,
then $X$ is a standard \sy , and a complete reduction is obtained:
\[ \dot r=h_1(r,z)r \q,\q \dot z=h_3(r,z)z \q,\q \dot\theta=-h_2(r,z)\]

\sk\ni
b)  only  $h_1$ and $h_3$ are functions of $r$ and $z$,
then $X$ is a $\la$-\sy , and $\dot r$ and $\dot z$ are as in a), but 
$\dot\theta=-h_2(r,z,\theta)$

\sk\ni
c)  $h_2/h_1$ and $h_3/h_1$ are functions of $r$ and $z$,
then $X$ is a orbital \sy\ and we have reduction up to a common factor
\[ \dot r=h_1(r,z,\theta)r \q,\q \dot z=h_1(r,z,\theta)\chi_a(r,z)z\q,\q  
\dot\theta=-h_1(r,z,\theta)\chi_b(r,z)\]
giving 
\[ \frac{\d r}{\d z}\=\Psi_1(r,z)\q,\q  \frac{\d \theta}{d r}\=\Psi_2(r,z)\]

\sk\ni
d)  only $h_3/h_1$ is function  of $r$ and $z$,
then $X$ is a orbital $\si$-\sy, and $\dot r$ and $\dot z$ are as in 
c), but $\dot\theta=-h_2(r,z,\theta)$.

\sk\sk
Another related result, concerning the presence of constants of 
motions of the DS having the property 
of being simultaneously invariant under the \sy\ is the following:

\sk\ni
{\bf Corollary.} \emph{In the above hypotheses, if a DS admits 
a rank $r$ involutive set of $\si$-\sys , or orbital $\si$-\sys , there 
are $n-r-1$ constants of motion, independent of time,  of the DS, which 
are also invariant under all the $\si$-\sys\ $X_\al$. }

\sk
This is obtained (using again  Frobenius theorem) looking for common 
invariants of the extended $(s+1)$-dimensional set 
$\widehat{\X}:=\{F,X_\al\}$ (with $F=f\cdot\nabla_u$ as before),
or $\widehat{\phi}:=\{f,\phi_\al\}$.  
An extension to 
non-autonomous DS and time dependent constants of motion can be easily 
obtained.

\section{From DS to higher-order ODE's}

Any ODE $\E\big(u(t)\big)=0$ of order $n>1$ can be transformed into a DS, as well 
known.
Writing $u^{(n)}=p(t,u,\dot u,\ldots)$, if the ODE does not contain 
explicitly the independent variable $t$, then one can put as usual
\[u=u_1\ ,\   \dot u_1=u_2\ ,\ldots, \   \dot u_n=p(u,\dot u,\ldots)\]
if instead $p$ depends on $t$, one simply includes the new variable 
$u_0=t$ and the \eq\ $\dot u_0=1$.
The converse is ``in principle" (locally, and apart from degenerate cases) also 
true (see \cite{NuLe,CGW}), but the transformation of a DS into an ODE 
requires the inversion of some implicit expressions.

Let us show the procedure in the case of a DS with 3 dependent variables 
$u^a$. If the DS is autonomous, $\dot u^a=f^a(u)$, then one puts
\[u_1:=y_1:=y\  ,\  \dot u_1=f_1(u):=y_2=\dot y, \]
\[\dot y_2=D_t f_1(u)=f\cdot\nabla_u\,f_1(u):=\Phi(u):= y_3=\ddot y\]
then one has to express $u_2$ and $u_3$ in terms of $y_1,y_2,y_3$ using 
the two above definitions, and finally one gets
\[\dot y_3=\   \stackrel{\ldots}{y}=D_t\Phi\big(u(y)\big):=p(y)\]
which produces the ODE
\[   \stackrel{\ldots}{y}\=p(y,\dot y,\ddot y)\ . \]
If the DS is non-autonomous, then it can be ``autonomized'' introducing as 
usual $u_0=t$, and the above procedure can be adapted accordingly.

The procedure of transforming a DS into a higher-order ODE opens 
interesting possibilities of reducing the ODE. If indeed the DS admits 
some \sy\ (including $\si$-\sys\ and orbital \sys ), then we have shown
that  the DS can be reduced in terms of suitable \sy -adapted variables.
This  reduction is immediately transferred, up to the change of variables
described  above, to the higher-order ODE. Observing that not all \sys\
of the DS  become
automatically Lie point-\sys\ of the ODE, we get a sort of ``reduction of 
the ODE's without \sys ".
There are several possibilities in this direction, as we will show in the 
following examples.

\sk

To illustrate the procedure, we give first an example of  DS admitting a 
standard \sy ; we   construct the corresponding higher order ODE and 
show how the \sy\ property of the DS can be used to obtain a reduced \eq\ 
for the ODE. In this example, the reduced \eq\ can be easily solved and this 
procedure provides thus an alternative way to get the full \so\ of the ODE.
Examples 6 and 7 deal with ODE's deduced from DS admitting resp. a
$\la$-\sy\ and a $\si$-\sy .

\bk\ni
{\it Example 5.} The DS
\[\dot u_1=u_1+u_1^2u_2\q,\q \dot u_2=u_2+u_1u_2^2\]
admits the standard \sy\
%\[X\=u_1\frac{\pd}{\pd u_1}-u_2\frac{\pd}{\pd u_2}\]
$X=u_1\pd/\pd u_1-u_2\pd/\pd u_2$.
An invariant variable under $X$ is $w=u_1u_2$, which satisfies the reduced \eq\ 
$\dot w=2w+2w^2$. The ODE obtained through the positions $u_1=y,\,\dot 
u_1=y_2=\dot y$ etc. is
\[\ddot y\=-2\dot y+3\frac{\dot y^2}{y}\ .\]
Integrating the reduced \eq\ for $w$ and passing to the new variable $y$
we obtain the reduced \eq\  for the ODE
\[\frac{\dot y}{y}\=\frac{c\, \exp(2t)}{1-c\,\exp(2t)}-1\]
and from this the full \so\ of the ODE
\[y\=\frac{c'\exp(t)}{\sqrt{c\,\exp(2t)-1}}  \]
where $c,c'$ are constants. \EOE

\bk\ni
{\it Example 6.} This is a simple example with a $\la$-\sy . We start
from the  DS
\[ \dot u_1\=u_2 \q,\q \dot u_2\=2u_2^2/u_1 \]
having a (standard) dilation \sy\ $X=u_1\pd/\pd u_1+u_2\pd/\pd u_2$.
Using Prop. 1 with $\mu=u_1$, the new DS
\beq\lb{exb} \dot u_1\=u_2+u_1^2 \q,\q \dot u_2\=2u_2^2/u_1+u_1u_2 \eeq
admits the above $X$ as a $\la$-\sy . With $u_1=y,\,u_2=\dot y-y^2$, 
according to the above described procedure, we get the ODE 
\[\ddot y\=2\frac{\dot y^2}{y}-y\dot y+y^3 \ .\]
The DS (\ref{exb}) can be reduced by the $X$-invariant variable 
$w=u_2/u_1$, indeed $\dot w=w^2$; the same reduction holds for the new 
variable $\~w=(\dot y/y)-y$, as easily checked. The reduced 
\eq\ for $w$ can be immediately solved producing the 
(time-dependent) first integral for the ODE $\kappa=t+y/(\dot y-y^2)$ = const. 
This   \eq\ for $y(t)$ can be  further integrated 
giving the general \so\ of the ODE
\[y(t)\=\ \Big((c-t)(c'+\log(c-t)\Big)^{-1} \]
where $c,c'$ are constants.
As above, the \so\ of the ODE could be obtained
(although not too easily) also by standard methods, but this example  can
be useful to further illustrate this
\sy-based procedure. \EOE

\sk\sk\ni
{\it Example 7.} This is an example where an ODE is constructed starting 
from a DS admitting a $\si$-\sy . The very simple DS
\[\dot u_1=1\q,\q \dot u_2=u_3\q ,\q \dot u_3=u_2\]
admits the two standard \sys
\[X_1\=\frac{\pd}{\pd u_1} \q ,\q X_2\=u_2\frac{\pd}{\pd u_2}
+u_3\frac{\pd}{\pd u_3}\ .\] 
Using Prop.1 with $\mu_1=u_3,\,\mu_2=1/u_1$, we
obtain the new DS
\beq\lb{exc}\dot u_1=1+u_3\q,\q \dot u_2=u_3+u_2/u_1\q,\q 
\dot u_3=u_2+u_3/u_1\eeq
which then admits the two \vf s $X_1,X_2$ as $\si$-\sy . A common invariant 
under these \vf s is $w=u_2/u_3$ which satisfies the \eq\ $\dot w=1-w^2$.
The ODE which is deduced from the above DS (\ref{exc}) is
\[  \stackrel{\ldots}{y}\=\dot y-1+2\frac{\ddot y}{y}+\frac{(\dot y-1)^2}{y^2}\ .\]
After integration of the  \eq\ for $w$, passing to the new 
variable $y$, one obtains the reduced \eq\ for $y(t)$
\[\frac{y\ddot y-\dot y+1}{y(\dot y-1)}=\frac{\exp(2t)-c}{\exp(2t)+c} \]
where $c$ is a constant. \EOE

\sk\sk
The two next and final examples deal with the case of $\si$-orbital \sys .
According to Prop.2, we can construct orbitally symmetric DS starting 
from a $\si$-symmetric (or standardly symmetric as well) by 
multiplication by an arbitrary function $\rho(u)$. A good choice for this 
function may be, e.g., $\rho=1/f_1(u)$ with usual notations, in such a 
way that the new DS becomes, renaming for convenience the variables 
$u_1,\ldots,u_n$ as $v_0,\ldots,v_{n-1}$
\beq\lb{uv}\dot v_0=1\q,\q \dot v_1=f_2(v)/f_1(v)\q,\ldots,\q\dot 
v_{n-1}=f_n(v)/f_1(v)\eeq
i.e. in the form of an ``autonomized'' DS where $v_0=t$ and then $v_1=y$, 
$\dot v_1=f_2(v)/f_1(v)=\dot y$, $\dot v_2=D_t\big(f_2(v)/f_1(v)\big)=\ddot y$, etc. 
The ODE deduced 
from this DS will then be of order $(n-1)$. We shall adopt this choice 
for the function $\rho(u)$  in both the following examples.

As said above, in the case of orbital \sys , we need {\it at least  two} 
invariants $w_j$ under the \vf s $X_\al$ in order to have a reduced \eq .
This may be reached either with two invariants under a single \vf\ (hence in the 
case of a single standard \sy , or also a $\la$-\sy\ as considered in Example 8 below), 
or with two common invariants under two \vf s (standard, or also $\si$-\sy\ as 
 in Example 9).

\sk\sk\ni
{\it Example 8.} The DS
\[\dot u_1\=u_1u_2\q,\q\dot u_2\=u_1/u_3\q,\q\dot u_3=u_3\]
admits the standard \sy\
%\[X\=u_1\frac{\pd}{\pd u_1}+u_3\frac{\pd}{\pd u_3}\]
$X\=u_1 \pd/\pd u_1+u_3 \pd/\pd u_3$.
Using Prop.1 with $\mu=u_2$ and then Prop.2 with $\rho=1/(u_1u_2)$ 
we   get the DS, using the notations introduced in (\ref{uv}),
\[
\dot v_0\=1\q,\q
\dot v_1\=\frac{1}{2v_1 v_2}\q,\q 
\dot v_2\=v_2\frac{1+v_1}{2v_0 v_1}
\]
which admits the above \vf\ $X$ as an orbital $\si$- (actually, a $\la$) 
-\sy . The ODE which  can be deduced from this DS is
\[\ddot y\=-\frac{\dot y(2t\dot y+y+1)}{2ty}\ .\]
There are two 
independent invariants under the above \vf\ $X$, namely 
$w_1=u_2=v_1,\,w_2=u_1/u_3=v_0/v_2$; they satisfy the \eq s
\[\dot w_1=\frac{1}{v_0}\frac{w_2}{2w_1} \q,\q 
\dot w_2=\frac{1}{v_0}w_2\frac{w_1-1}{2w_1}\]
and from these we obtain the reduced \eq\ $\d w_2/\d w_1=w_1-1$ which 
can be easily integrated giving in the new variable $y$ the reduced \eq\ 
for $y(t)$ (and a first integral for the ODE)
\[2ty\dot y-\frac{1}{2}y^2+y\= {\rm const}\ .\]

\sk\ni
{\it Example 9.} The simple DS
\[\dot u_1=0\q,\q \dot u_2=u_3\q,\q \dot u_3=u_4 \q,\q \dot u_4=u_2\]
admits the two standard \sys
\[X_1\=\frac{\pd}{\pd u_1}\q,\q X_2\=u_2\frac{\pd}{\pd
u_2}+u_3\frac{\pd}{\pd u_3}+u_4 \frac{\pd}{\pd u_4}\ .\]
Thanks to Prop.1 and 2, with $\mu_1=u_2,\,\mu_2=u_1$ and $\rho=1/u_2$
we obtain the DS, with the notations as in (\ref{uv})
\[\dot v_0=1\q,\q \dot v_1=v_0+v_2/v_1\q,\q \dot v_2=
(v_3+v_0v_2)/v_1\q,\q\dot v_3=1+v_0v_3/v_1\ .\]
Two independent common invariants under $X_1,X_2$ are
$w_1=u_2/u_3=v_1/v_2,\,w_2=u_4/u_3=v_3/v_2$. 
The corresponding ODE is 
\[  \stackrel{\ldots}{y}=\frac{1}{y^2}\Big(1-(\dot y-t)^3-ty+3ty\ddot y-4y\dot y\ddot 
y\Big)\]
which admits the reduced \eq
\[\frac{\d w_1}{\d w_2}\=\frac{1-w_1w_2}{w_1-w_2^2}\ .\]

\end{document}